\documentclass[aps,prl,twocolumn,groupedaddress,showpacs]{revtex4}
\usepackage{epsfig}
\usepackage{graphicx}
\usepackage{amsfonts}
\usepackage{latexsym}
\usepackage{bm}

\def\bk{{\bf k}}
\def\bx{{\bf x}}
\def\by{{\bf y}}

\def\8{\infty}

\def\d{\partial}

\def\undertext#1{\vtop{\hbox{#1}\kern 1pt \hrule}}

\def\VEV#1{\left\langle\,#1\,\right\rangle}

\def\be{\begin{equation}}
\def\ee{\end{equation}}
\def\bea{\begin{eqnarray} & &}
\def\eea{\end{eqnarray}}

\def\rf#1{(\ref{#1})}

\def\rf#1{(\ref{#1})}
\def\t{\tilde}

\def\rfs#1{Eq.~\rf{#1}}

\begin{document}


\title{Phonons in Random Elastic Media and the Boson Peak}


\author{V. Gurarie}
\affiliation{Department of Physics, CB390, University of Colorado,
Boulder CO 80309, USA}
\author{A. Altland}
\affiliation{Institut f\"ur Theoretische Physik, Universit\"at zu
K\"oln, Z\"ulpicher Str 77, K\"oln, Germany}


\date{\today}

\begin{abstract}
We show that the density of states of random wave equations,
normalized by the square of the frequency,  has a peak ---
sometimes narrow and sometimes broad --- in the range of wave
vectors between the disorder correlation length and the
interatomic spacing. The results of this letter may be relevant
for understanding vibrational spectra and light propagation in
disordered solids.
\end{abstract}


\maketitle

One of the intriguing features of random elastic media ---
observable in both Raman or neutron cross sections but also in
calorimetric measurements --- is an anomalous accumulation of
phonons at low frequencies \cite{Jackle1981}. This phenomenon
finds its most prominent manifestation in a peak in the quantity
$I(\omega)=\rho(\omega)/\omega^2$, where $\rho(\omega)$ is the
phonon density of states (DoS) and $\omega$ is the frequency. In
recent years, many competing theories as to the origin of this
so--called boson peak have been formulated (for a recent
reference, see Ref.~\cite{Parshin2003} and references therein). Of
these approaches a majority is based on model mechanisms specific
to the low temperature physics of amorphous solids. Yet the boson
peak is shown by both glassy and random crystalline systems alike,
an observation which has ignited the search for an explanation
which is not tied to the specifics of a glassy environment.

\begin{figure}[htbp]
\centerline{\resizebox{3.5in}{!}{\includegraphics{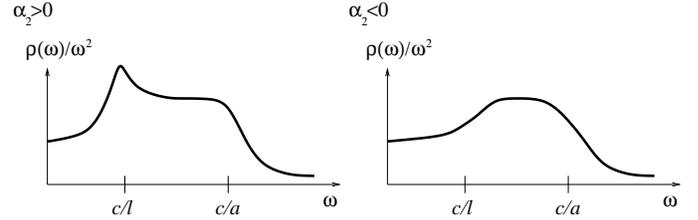}} }
\caption{\label{graph1} $I(\omega)$ for the random elastic media
in cases where $\alpha_2>0$ or $\alpha_2<0$.}
\end{figure}

On a basic level, the acoustic excitations of both amorphous
materials and disordered crystals are described by random wave
equations. Existing analyses of such equations in the literature
indeed predicted a disorder generated excess DoS,
\cite{Schirmacher1998}. However, these structures where observed
at high frequencies (wavelengths of the order of the interatomic
spacing), while the boson peak is a low energy phenomenon. In this
letter we argue that the DoS of elastic vibrations in disordered
media is enhanced by disorder at low frequencies, with $I(\omega)$
exhibiting either a peak or a broad maximum (see Fig.~\ref{graph1}
for a representative picture of the DoS). Which of the two will be
observed crucially depends on the modelling of the randomness, i.
e. the \textit{type} of disorder at work. Since the profile of the
randomness of `real' systems is generally unknown, we are not in a
position to judge whether the present mechanism alone may account
for the spectral peaks observed in all experiments. We believe,
however, that it operates under quite general conditions.


Let us begin by discussing some generic large scale structures of
the DoS of acoustic excitations. In a clean system, the linear
relation $\omega = \bar c k$ between frequency and magnitude $k$
of the wave vector gives rise to a DoS
\begin{equation}
\label{eq:naiveDoS} \rho(\omega) = \int {d^d k \over (2 \pi)^d
}~\delta(\omega-\bar c k) = {A_d \over (2 \pi \bar c)^d }~
\omega^{d-1}.
\end{equation}
Here, $A_d$ is the area of the $d$--dimensional unit sphere and
$\bar c$ the speed of sound (for our present analysis, the
potential existence of several acoustic phonon branches with
different sound velocities will not be of importance). Tied to the
linearizability of the dispersion relation, the relation $\rho
\sim \omega^{d-1}$ applies only to  frequencies $\omega\ll
\omega_D$ much smaller than the Debye frequency $\omega_D\sim
{\bar c\over a}$ ($a$: lattice spacing.) On dimensional grounds,
the generalization of (\ref{eq:naiveDoS}) to higher frequencies
must be of the form
\begin{equation}
\label{eq:latt} \rho(\omega) \sim \omega^{d-1} \tilde
f\left(\omega/
    \omega_D\right),
\end{equation}
where $\tilde f(u)$ is some function with $\tilde f(u\ll 1)\simeq
1$. It is clear  that $\t f(u)$ falls off to zero at $u \gg 1$.

In contrast, in disordered materials we find strong deviations
from the above scaling form already for frequencies $\omega\ll
\omega_D$. Indeed, our analysis of the random wave equation below
will lead us to
\begin{equation}
  \label{eq:DoSscalling}
  \rho(\omega)= {A_d \over (2 \pi \bar c)^d }~\omega^{d-1}
  f\left({\omega l\over \bar c}\right),\;\;
  \omega \ll \omega_D,
\end{equation}
where the effective correlation length $l\gg a$ of the disorder is
assumed to be much larger than the lattice spacing, and $\bar c$ is
the typical speed of sound in random media to be defined more
precisely below. The scaling function $f$ is given by
\begin{eqnarray}
\label{eq:fexp}
f(u) &=& 1 + \alpha_1 u^2 + \dots, \ u \ll 1, \cr
f(u) &=& \beta \left( 1 + {\alpha_2 \over u^2} + \dots \right),
\;1\ll  u \ll {l \over a},
\end{eqnarray}
where the form of the coefficients $\alpha_{1,2}$, $\beta$ depends
on the space dimensionality and on a few basic characteristics of
the randomness. In the most interesting case, $d=3$, we find that
$\beta>1$ and $\alpha_1>0$. The sign of $\alpha_2$ lacks
universality. $\alpha_2$ is negative if disorder in materials is
mostly in the random elastic constants, and it is positive if
disorder is mostly due to fluctuating mass density.

For $\alpha_2>0$, $I(\omega)$ grows for $\omega \ll \bar c/l$ and
falls off at $\omega \gg \bar c/l$ (while still $\omega \ll
\omega_D$). This implies the existence a peak
at $\omega \sim \bar c/l$. In contrast, for $\alpha_2<0$,
$I(\omega)$ increases until $\omega$ becomes much larger than
$\bar c/l$.  Combined with the expected drop--off of $I(\omega)$
at $\omega \gtrsim \omega_D$, this produces a broad maximum for
$I(\omega)$ between $\bar c/l$ and $\omega_D$. Fig.~\ref{graph1}
shows a caricature of the two scenarios. We finally
note that if the disorder is relatively weak, ${\rm rms\,}c<\bar
c$, (so that the fluctuations of local speed of sound $c$ are
smaller than the typical sound velocity) $\alpha_1\sim
\alpha_2\sim {\,\rm var}\, c/\bar c^2$; for more generally
applicable expressions, see below.



We next turn to the derivation of these results. Consider the
wave equation
\begin{equation}
  \label{eq:random_wave}
  \left(\Delta + {\omega^2 \over c^2({\bf x})}\right) \psi({\bf x})=0,
\end{equation}
where the random velocity field $c({\bf x})$ fluctuates on spatial
scales $\sim l$.  Interpreting the variable $m({\bf x})\equiv
1/c^2({\bf x})$ as the mass density of the random medium, we refer to
\rfs{eq:random_wave} as a `random mass' wave equation.

In the regime of small frequencies, $\omega \ll \bar c/l$, we are
facing a situation where the correlation length $l$ is much smaller
than the `typical' wave length $k^{-1} \equiv \bar c/\omega$ at which
the wave function $\psi({\bf x})$ fluctuates.  The phonon field
effectively averages over many fluctuation intervals of the disorder
and, to a first approximation, \rfs{eq:random_wave} may be replaced
for its average over random $c({\bf x})$: $\left(\Delta + {\omega^2
\over \bar c^2}\right) \psi({\bf x})=0$, where the typical speed of
sound $\bar c$ is defined as
\begin{equation}
\label{eq:typc} \bar c \equiv \langle c({\bf
x})^{-2}\rangle^{-1/2}
\end{equation}
and the angular brackets denote averaging over random $c({\bf x})$.
The very low frequency asymptotics is then given by \rfs{eq:naiveDoS}
with $\bar c$ taken from \rfs{eq:typc}.  (This result, as well as the
high frequency asymptotics \rfs{eq:beta} discussed below, are
originally due to Chalker \cite{ChalkerPrivate}.)

 To compute corrections to the low frequency asymptotics of the
DoS we rewrite Eq.~(\ref{eq:random_wave}) as
\begin{equation}
  \label{eq:rw_pert}
    \left(- \Delta - {\omega^2\over \bar c^2} h(\bx)\right) \psi({\bf x})= {\omega^2
    \over \bar c^2} \psi(\bx),
\end{equation}
where the function $ h({\bf x})\equiv \left[{\bar c/ c({\bf
x})}\right]^2-1 $ describes the randomness. This representation
suggests to interpret $\omega^2/\bar c^2$ as the eigenvalue of the
operator $\Delta$ weakly perturbed by $\omega^2 h(\bx)/\bar c^2$.
The unperturbed problem is trivially
diagonalized by a set of plane waves $ \psi_\bk(\bx) \equiv
{1\over L^{d/2}} \exp(i\bk \cdot \bx), $ with eigenvalues
$\omega^2/\bar c^2 = k^2$ ($L$ is the system size.) We next apply
standard perturbation theory to compute the eigenvalue shift
caused by the presence of the perturbation $\omega^2 h/\bar c^2$.
The vanishing of $\VEV{h}$ implies that, on average, there are no
first order corrections. To second order in $h$ we find that the
average eigenvalue is given by 
$
  {\omega^2\over \bar c^2} = k^2 + k^4 \int {d^d
k' \over (2 \pi)^d} {g({\bf k'}) \over k^2-k'^2}$,
where $g({\bf k})$ is the Fourier transform of the disorder
correlation function of $g({\bf x}-{\bf
y})\equiv \VEV{h({\bf x}) h({\bf y})}$. Irrespective of the
distribution of the disorder, this function (a) drops off rapidly for
$k>l^{-1}$ and (b) approaches a constant value $\sim  l^d {\rm
\,var}(h)$ for $k\ll l^{-1}$. This implies that
for $d> 2$, the integral above is dominated by momenta $k'\sim l^{-1}$.
At the same time, the reference momentum $k\sim \omega/\bar c\ll
l^{-1}$ is small. We thus neglect the $k$--dependence of the
integrand and arrive at
\begin{equation}
  \label{eq:eval_lowo}
  {\omega^2\over \bar c^2} = k^2 -  k^4 \int {d^d
k' \over (2 \pi)^d} {g({\bf k'}) \over k'^2}.
\end{equation}
Physically, the correction to the
zeroth order eigenvalue is due to virtual scattering events wherein
states with low--lying momentum $k\sim \omega/\bar c$ scatter off
rapid fluctuations of $h$ into high--lying states $k'\sim l^{-1}$.
Yet for small frequencies, the large phase volume $\sim l^{-d}$
available to these scattering processes cannot outweigh the overall
multiplicative factor $\omega^4$.  This mechanism pervades to higher
orders in the eigenvalue expansion and justifies the perturbative
approach.  In particular, Eq.  (\ref{eq:eval_lowo}) indeed describes
the dominant correction to the low frequency dispersion relation.

We
next substitute Eq.  (\ref{eq:eval_lowo}) into $\rho(\omega) = {1\over
L^d} \sum_\bk \delta(\omega(k)-\omega)$ to obtain an expansion of the
DoS as in Eqs.  (\ref{eq:DoSscalling},\ref{eq:fexp}).  Specifically,
the coefficient
\begin{equation} \label{eq:alpha1} \alpha_1 ={d+2
\over 2l^2} \int {d^d k \over (2 \pi)^d} {g({\bf k}) \over k^2}\sim
\mathrm{var}(h)>0,
\end{equation}
where the proportionality to $\mathrm{var}(h)$ follows from the
properties of the correlation function $g$ discussed above.
Interestingly, positivity of $\alpha_1$ in \rfs{eq:alpha1} is a
direct consequence of the fact that the second order perturbation
theory always lowers the ground state energy.

For small velocity fluctuations, ${\rm \,var}(c) \ll \bar c^2$, we
obtain the estimate $\alpha_1\sim \mathrm{var}(h)\sim{\rm
var}(c)/\bar c^2$ quoted above. We finally note that these results
hold only for $d>2$. For $d \le 2$, the integral in
\rfs{eq:alpha1} is infrared divergent and the approximation scheme
employed here breaks down.


We next turn to the discussion of the high frequency case, $\omega
\gg \bar c/l$. In this regime, the velocity field varies very
little over length scales comparable to the typical wavelength.
This implies that, locally, the solutions of
(\ref{eq:random_wave}) behave like plane waves with the local
dispersion relation $k({\bf x})=\omega/c({\bf x})$ and the local
density of states  given by
$$
\rho(\omega,{\bf x}) = {A_d \over (2 \pi)^d c^d({\bf x})}~
\omega^{d-1}.
$$
Averaging this result over the random velocity field we obtain the
global density of states
$$
\rho(\omega) = {A_d \over (2 \pi)^d} \VEV{1 \over c^d({\bf
x})}~\omega^{d-1}, \qquad\omega \gg {\bar c \over l}.
$$
We thus find that the coefficient $\beta$, introduced in
\rfs{eq:fexp}, is given by
\begin{equation}
\label{eq:beta} \beta= {\VEV{c^{-d} \left( {\bf x} \right)} \over
\VEV{c^{-2}({\bf x})}^{d \over 2}}.
\end{equation}
As a consequence of the convexity of the power function, $\beta>1$
for $d>2$.

 To obtain corrections to the high frequency
 asymptotics~(\ref{eq:beta}), we
 need
to compute distortions in the spectral density caused by shallow
(compared to the wave length) variations  of the velocity field.
It turns out that this task is most efficiently tackled by
analyzing the Green's function
\begin{equation}
\label{eq:greens} G=-\left( {1\over \omega^2} \nabla^2 + m(\bx) +
i \epsilon \right)^{-1},
\end{equation}
where $m\equiv c^{-2}$. From (\ref{eq:greens}) the
average DoS is obtained as
\begin{equation}
\label{eq:dosgr} \rho(\omega) =  {2  \over \pi \omega} ~{\rm Im}
\left\langle m({\bf x})~G(\bx,\bx)\right\rangle,
\end{equation}
What makes the operator (\ref{eq:greens}) a good  starting point
for our analysis is its structural similarity to a Schr\"odinger
operator with `Planck's constant' $\hbar\sim \omega^{-1}$. This
analogy will enable us to apply semiclassical approximation
schemes familiar from quantum mechanics.  We begin by applying the
Wigner transform
$$
G(\bx,\by) = {\omega^d \over (2 \pi)^d} \int d^d k~G(\bx,\bk)~e^{i
\omega
  \bk\cdot \left(\bx-\by\right)},
$$
whereupon the `Schr\"odinger equation' assumes the form
\begin{equation}
\label{eq:Schro} \left[\left({\bf k}-i\omega^{-1}\partial_{\bf
x}\right)^2 -m({\bf x})-i\epsilon\right]G(\bx,\bk)=1.
\end{equation}
To make use of the smallness, $\sim \omega^{-1}$, of the
derivative operators, we expand $G(\bx,\bk)$ in powers of
$\omega^{-1}$,
$$
G({\bf x},{\bf k}) = G^{(0)}({\bf x},{\bf k})+ \frac{1}{\omega}
G^{(1)}({\bf x},{\bf k}) + \frac{1}{\omega^2} G^{(2)}({\bf x},{\bf
k})+ \dots,
$$
substitute the expansion into \rfs{eq:Schro}, and find $G^{(0)}$,
$G^{(1)}$, $G^{(2)}$, $\dots$, recursively. By symmetry,
$G^{(1)}$ vanishes upon configurational averaging so that the
dominant correction to the DoS is provided by $G^{(2)}$.
Substituting this coefficient into \rfs{eq:dosgr}, and comparing
with Eq.~(\ref{eq:fexp}) we obtain the result
$$
\alpha_2 ={\left(d-2\right)^2 \left(4-d \right) \over 24} { l^2
\over \bar c^2 \VEV{c^{-d}}}  \VEV{{ \nabla c \cdot \nabla c \over
c^d}}.
$$
Clearly, $\alpha_2>0$ if $d=3$. If the fluctuations of $c$ are weak,
$\alpha_2 \sim {\rm var}\,c/\bar
c^2\sim \alpha_1$.

We thus derived \rfs{eq:fexp}. The positivity of the two expansion
coefficients $\alpha_{1,2}$ implies that the scaling function
$f(u)$ in \rfs{eq:DoSscalling} grows at $u \ll 1$ and falls off at
$u \gg 1$. This implies that the functional profile of the DoS
contains a peak somewhere at $u =\mathcal{O}(1)$.

To conclude our analysis of the prototypical wave equation
(\ref{eq:random_wave}), let us briefly discuss the case of space
dimensions $d\not=3$. For $d=1$, the high frequency coefficient
$\alpha_2>0$, yet $\beta<1$. Due to the instability in the the low
frequency expansion, the methods applied in this letter do not
allow to calculate $\alpha_1$ at $d \le 2$. However, for certain
distributions of the disorder transfer matrix methods may be
applied to obtain an exact solution~\cite{Gurarie2002a}. These
calculations show that $f(u)$ is a globally decreasing function,
and $\alpha_1<0$. For $d=2$, the two dominant high--frequency
coefficients are structureless, $\beta=1$ and $\alpha_2=0$, while
the low frequency expansion in \rfs{eq:fexp} will now involve
terms proportional to $\log(u)$. From these results we cannot
decide whether $\rho(\omega)/\omega$ has a maximum or a minimum
between its low and high frequency asymptotics. Finally, for
$d>3$, $\alpha_1>0$ and $\beta>1$, however $\alpha_2\le 0$, which
implies that $f(u)$ is a monotonously increasing function.
Summarizing, we see that  the normalized DoS of
\rfs{eq:random_wave}, $\rho(\omega)/\omega^{d-1}$, exhibits a peak
at wavelengths of the order of disorder correlation length only
for $d=3$.

Before proceeding, let us briefly compare our so far results to
earlier work.  In most numerical simulations of the problem (cf.
e.g., Ref.~\cite{Schirmacher1998} and references therein), the
disorder is chosen to be uncorrelated, that is, its correlation
length is of the order of the lattice spacing.  For such type of
disorder, $u\simeq 1$ translates to $\omega\sim \omega_D$ deep in
the bulk of the spectrum.  At these frequencies it is hard to tell
whether deviations from the low frequency asymptotics
$\rho(\omega)\sim \omega^2$ are caused by lattice effects (cf.
Eq.~(\ref{eq:latt})) or by disorder (Eq.~(\ref{eq:DoSscalling})).
Qualitatively, however, the numerical data is in agreement with
the results of our present analysis.  Turning to earlier
analytical work, we notice that most approaches to random elastic
problems rely on the self consistent Born approximation (SCBA).
(See Ref.~\cite{EfetovBook} for a general review of the methods
involved and Refs.~\cite{John1982,Schirmacher2002}  for the
applications of these methods to random wave equations.)  However,
for a number of reasons the results obtained by this method lack
quantitative reliability when used to calculate the DoS: (i)
Within the standard SCBA, $1/c^2$ is modelled as a Gaussian
distributed variable. This implies the existence of rare domains
where $1/c^2$ is negative. Using the techniques introduced in
Ref.~\cite{Gurarie2003b} it is possible to show that these
domains, no matter how small, lead to unstable wave modes and to
unphysical sharply growing contributions to the density of states
at higher frequencies.
(ii) The SCBA neglects certain contributions (`crossed diagrams')
to the perturbative expansion of the Green function
(\ref{eq:greens}).  At high frequencies, this approximation
becomes invalid. The value of the threshold frequency beyond which
the SCBA breaks down depends on the type of disorder under
consideration. At any rate, however, it is parametrically smaller
than the band width. At the same time, (iii), previous studies
focused on the case of short range correlated disorder. In this
case, deviations of the scaling law $\rho\sim \omega^{d-1}$ are
expected at frequencies $\omega \sim \omega_D$, i.e. in a regime
where the SCBA breaks down {\it and} lattice effects intervene. (The alternative techniques in
this letter avoid all of these problems.)

The wave equation \rfs{eq:random_wave} applies to the specific
case, where only the mass density of the elastic medium
fluctuates. The continuum description of a more general
environment wherein the elastic constants also fluctuate reads as
\begin{equation}
\label{eq:bond1} \left( \nabla \mu({\bf x}) \nabla + \omega^2
m({\bf x}) \right) \psi({\bf x}) =0.
\end{equation}
Here both $\mu({\bf x})$ and $m({\bf x})$ are random positive
quantities.

It turns out that the direct perturbative expansion applied above
to \rfs{eq:random_wave} cannot be used to determine the
coefficient $\alpha_1$ of the problem \rfs{eq:bond1}. As an
alternative, we apply a generalized variant of the
self--consistent Born approximation, wherein
$\mu(\bx)=\mu_0+\sigma^2(\bx)$, and $\sigma(\bx)$ is a Gaussian
distributed variable with zero mean. (In this way, positivity of
the elastic constant is ensured.) The actual implementation of the
SCBA for this type of disorder turns out to be rather involved and
its details will be published elsewhere. Suffice to say that at
$d=3$, $\alpha_1$ is still positive. $\alpha_2$, on the other
hand, no longer has definite sign. Applying the high frequency
expansion outlined above for the random mass density case we find
\begin{eqnarray}
\label{eq:bond} \alpha_2 ={l^2 \over \bar c^2 \VEV{m^{3 \over 2}
\mu^{-{3 \over 2}}}} \left[{1 \over 96} \VEV{ {\mu^2 \left( \nabla
m \right)^2
\over  m^{3 \over 2} \mu^{5 \over 2}}} + \right. \\
\nonumber \left. {5 \over 48} \VEV{m \mu \nabla m \nabla \mu \over
m^{3 \over 2} \mu^{5 \over 2}} -{23 \over 96}\VEV {m^2
\left(\nabla \mu \right)^2 \over m^{3 \over 2} \mu^{5 \over
2}}\right].
\end{eqnarray}
If $\alpha_2$ is negative (which happens, for example, in the
limiting case of non--random $m({\bf x})$) $f(u)$ no longer has a
maximum at $u \simeq 1$. Instead, $I(\omega)$ is an increasing
function of $\omega$ with a broad maximum reached at frequencies
$\omega > \bar c/l$, before dropping off at frequencies higher
than $\omega_D$. Concluding, we find that, depending on the
profile of the disorder,
 the envelope function $I(\omega)$ of the random mass/elastic constants problem
 either contains a low--frequency peak, or a broad high frequency
 maximum.

So far, we have considered the case of scalar phonons.
In a realistic environment, however, $\psi\to u_i$ will be a
  $d$--component {\it vector}. The most general random vector problem
  would be governed by a formidable rank four random elastic modulus
  tensor. Assuming, however, a medium consisting of a
  random accumulation of `micro--crystallites' each of which
  possessing intrinsic rotational invariance --- this assumption may
  well be violated, especially in glassy environments, although it seems
  to work in polycrystalline materials --- the
  effective wave equation reduces to
\begin{equation}
\label{eq:elastic} \d_i \left[\lambda({\bf x}) \d_j u_j\right] +
\d_j \left[ \mu({\bf x}) \left( \d_i u_j + \d_j u_i \right)\right]
+ \omega^2 m({\bf x}) u_i =0,
\end{equation}
where $m(\bx)$ is a random density of the medium and $\lambda(\bx)$,
$\mu(\bx)$ are random Lam\'e coefficients.

In the limit where only $\lambda(\bx)$ is random, \rfs{eq:elastic}
can be mapped into \rfs{eq:random_wave} by substitution of
$\psi=\d_i u_i (\lambda + 2 \mu)$. In this case, both $\alpha_1$
and $\alpha_2$ are positive.
This result carries over to the case where, in addition to random
$\lambda(\bx)$, also the density $m(\bx)$ is
made random; For these two types of disorder
$I(\omega)$ has a peak at $\omega \sim \bar c/l$.
If, however, the shear modulus $\mu$ is also random, \rfs{eq:elastic}
becomes more similar to \rfs{eq:bond1}. Applying the SCBA, it is still
possible to show that $\alpha_1>0$ (for $d>2$). As with our previous
discussion of (\ref{eq:bond1}), however, $\alpha_2$ is no longer of
definite sign. Specifically, $I(\omega)$ will contain a broad maximum
($\alpha_2<0$) if
 only the shear modulus $\mu(\bx)$ is random.

 To conclude, we have shown that, depending on the type of disorder,
 the normalized DoS $I(\omega)$ of random wave equations  may either contain a peak at phonon
 wavelengths of the order of the disorder correlation length, or a
 broad maximum at wavelengths below the correlation length. We believe
 that this work is not only relevant for the interpretation of data on
 vibrational modes in random media but also to the analysis
 of other types of waves, such as electromagnetic waves in random
 environments.

 The authors are grateful to J. T. Chalker for many valuable
 discussions. VG is also grateful to A. V. Andreev for useful
 comments. Work supported by SFB/TR 12 of the Deutsche Forschungsgemeinschaft.

\bibliography{ref}

\begin{thebibliography}{9}
\expandafter\ifx\csname natexlab\endcsname\relax\def\natexlab#1{#1}\fi
\expandafter\ifx\csname bibnamefont\endcsname\relax
  \def\bibnamefont#1{#1}\fi
\expandafter\ifx\csname bibfnamefont\endcsname\relax
  \def\bibfnamefont#1{#1}\fi
\expandafter\ifx\csname citenamefont\endcsname\relax
  \def\citenamefont#1{#1}\fi
\expandafter\ifx\csname url\endcsname\relax
  \def\url#1{\texttt{#1}}\fi
\expandafter\ifx\csname urlprefix\endcsname\relax\def\urlprefix{URL }\fi
\providecommand{\bibinfo}[2]{#2}
\providecommand{\eprint}[2][]{\url{#2}}

\bibitem[{\citenamefont{J\"ackle}(1981)}]{Jackle1981}
\bibinfo{author}{\bibfnamefont{J.}~\bibnamefont{J\"ackle}}, in
  \emph{\bibinfo{booktitle}{Amorphous Solids}}, edited by
  \bibinfo{editor}{\bibfnamefont{W.~A.} \bibnamefont{Phillips}}
  (\bibinfo{publisher}{Springer-Verlag}, \bibinfo{address}{Heidelberg},
  \bibinfo{year}{1981}), chap.~\bibinfo{chapter}{8}, p. \bibinfo{pages}{135}.

\bibitem[{\citenamefont{{Gurevich {\it et al.}}}(2003)}]{Parshin2003}
\bibinfo{author}{\bibfnamefont{V.~L.} \bibnamefont{{Gurevich {\it et al.}}}},
  \bibinfo{journal}{Phys. Rev. B} \textbf{\bibinfo{volume}{67}},
  \bibinfo{pages}{094203} (\bibinfo{year}{2003}).

\bibitem[{\citenamefont{{Schirmacher {\it et al.}}}(1998)}]{Schirmacher1998}
\bibinfo{author}{\bibfnamefont{W.}~\bibnamefont{{Schirmacher {\it et al.}}}},
  \bibinfo{journal}{Phys. Rev. Lett.} \textbf{\bibinfo{volume}{81}},
  \bibinfo{pages}{136} (\bibinfo{year}{1998}).

\bibitem[{\citenamefont{Chalker}()}]{ChalkerPrivate}
\bibinfo{author}{\bibfnamefont{J.~T.} \bibnamefont{Chalker}}, \eprint{private
  communication}.

\bibitem[{\citenamefont{Gurarie}(2002)}]{Gurarie2002a}
\bibinfo{author}{\bibfnamefont{V.}~\bibnamefont{Gurarie}},
  \bibinfo{journal}{unpublished}  (\bibinfo{year}{2002}).

\bibitem[{\citenamefont{Efetov}(1997)}]{EfetovBook}
\bibinfo{author}{\bibfnamefont{K.~B.} \bibnamefont{Efetov}},
  \emph{\bibinfo{title}{Supersymmetry in Disorder and Chaos}}
  (\bibinfo{publisher}{Cambridge University Press},
  \bibinfo{address}{Cambridge, UK}, \bibinfo{year}{1997}).

\bibitem[{\citenamefont{{John {\it et al.}}}(1982)}]{John1982}
\bibinfo{author}{\bibfnamefont{S.}~\bibnamefont{{John {\it et al.}}}},
  \bibinfo{journal}{Phys. Rev. B} \textbf{\bibinfo{volume}{27}},
  \bibinfo{pages}{5592} (\bibinfo{year}{1982}).

\bibitem[{\citenamefont{{Schirmacher {\it et al.}}}(2002)}]{Schirmacher2002}
\bibinfo{author}{\bibfnamefont{W.}~\bibnamefont{{Schirmacher {\it et al.}}}},
  \bibinfo{journal}{Phys. Stat. Sol. (b)} \textbf{\bibinfo{volume}{230}},
  \bibinfo{pages}{31} (\bibinfo{year}{2002}).

\bibitem[{\citenamefont{Gurarie and Altland}(2004)}]{Gurarie2003b}
\bibinfo{author}{\bibfnamefont{V.}~\bibnamefont{Gurarie}} \bibnamefont{and}
  \bibinfo{author}{\bibfnamefont{A.}~\bibnamefont{Altland}},
  \bibinfo{journal}{J Phys A: Math. Gen.} \textbf{\bibinfo{volume}{37}},
  \bibinfo{pages}{9357} (\bibinfo{year}{2004}).

\end{thebibliography}

\end{document}